\documentclass[9pt,conference]{IEEEtran}
\IEEEoverridecommandlockouts

\usepackage{cite}
\usepackage{amsmath,amssymb,amsfonts}
\usepackage{algorithmic}
\usepackage{graphicx}
\usepackage{textcomp}
\usepackage{xcolor}

\usepackage{url}
\usepackage{hyperref}
\usepackage{booktabs}
\usepackage{multirow}
\usepackage{tikz}

\def\BibTeX{{\rm B\kern-.05em{\sc i\kern-.025em b}\kern-.08em
    T\kern-.1667em\lower.7ex\hbox{E}\kern-.125emX}}

\begin{document}

\title{Semi-Supervised Contrastive Learning for Controllable Video-to-Music Retrieval
}

\author{Shanti Stewart$ ^{1, {\star}} $ \qquad Gouthaman KV$ ^{1, {\dagger}} $ \qquad Lie Lu$ ^{1, {\dagger}} $ \qquad Andrea Fanelli$ ^{1} $ \\

{$ ^{1} $ Dolby Laboratories}
}

\maketitle


\def\thefootnote{$ \star $}\footnotetext{Work done during an internship at Dolby Laboratories.}
\def\thefootnote{\arabic{footnote}}

\def\thefootnote{$ \dagger $}\footnotetext{These authors contributed equally to this work.}
\def\thefootnote{\arabic{footnote}}

\begin{abstract}

Content creators often use music to enhance their videos, from soundtracks in movies to background music in video blogs and social media content. However, identifying the best music for a video can be a difficult and time-consuming task. To address this challenge, we propose a novel framework for automatically retrieving a matching music clip for a given video, and vice versa. Our approach leverages annotated music labels, as well as the inherent artistic correspondence between visual and music elements. Distinct from previous cross-modal music retrieval works, our method combines both self-supervised and supervised training objectives. We use self-supervised and label-supervised contrastive learning to train a joint embedding space between music and video. We show the effectiveness of our approach by using music genre labels for the supervised training component, and our framework can be generalized to other music annotations (e.g., emotion, instrument, etc.). Furthermore, our method enables fine-grained control over how much the retrieval process focuses on self-supervised vs. label information at inference time. We evaluate the learned embeddings through a variety of video-to-music and music-to-video retrieval tasks. Our experiments show that the proposed approach successfully combines self-supervised and supervised objectives and is effective for controllable music-video retrieval.

\end{abstract}


\section{Introduction}
\label{sec:Introduction}

From movies to video blogs and social media content, music greatly enhances how we experience entertainment.
Music adds depth to a story, evoking strong emotions and enriching the narrative.
The synergy between the visuals and music is essential for impactful storytelling.
However, finding music that matches the style/genre/emotion of a video can be difficult and time-consuming.
Thus, a system that can automatically recommend music for a given video is very desirable.

Various approaches have been proposed for this task of retrieving music from video (and vice-versa).
Hong et al. \cite{Hong-208-CBVMR} train audiovisual embeddings via inter-modal ranking and intra-modal structure losses.
More recent works use self-supervised learning methods due to their widespread success.
In~\cite{Pretet-2021-design_choices}, the authors match short video segments to music segments based on aggregated video and audio embeddings.
However, this approach likely cannot capture nuanced audiovisual relationships that often require a broader temporal context \cite{Pretet-2021-language_of_music_video_clips}.
Suris et al. \cite{Suris-2022-MVPt} use self-supervised contrastive learning to model the artistic correspondence between music and video.
They use Transformer networks \cite{transformer} to model the temporal context and audiovisual alignment in long-form music videos.
McKee et al. \cite{McKee-2023-language_guided_mvr} extend the approach of \cite{Suris-2022-MVPt} to include languaged-guided video-to-music retrieval.

Aside from video-to-music retrieval, there are numerous works on music and audio retrieval from other modalities such as language \cite{Huang-2022-MuLan, Manco-2022-MusCALL, Wu-2023-CLAP}, speech \cite{Doh-2023-speech_to_music}, and visual queries \cite{Stewart-2024-EmoCLIM, Wilkins-2023-bridging}.

Although self-supervised approaches are effective in capturing the inherent audiovisual correspondence between music and video, incorporating supervised information can further enhance retrieval performance. While some studies \cite{Li-2019-query_by_video, Stewart-2024-EmoCLIM} have explored supervised learning for music and visual content, supervised methods for music-video retrieval\footnote{We use the term \emph{music-video retrieval} to refer to both video-to-music and music-to-video retrieval.} are largely under-explored.



Motivated by this, we propose \emph{Control-MVR}: a framework for \underline{Control}lable \underline{M}usic-\underline{V}ideo \underline{R}etrieval.
Our approach combines self-supervised and supervised training objectives, by leveraging both natural audiovisual correspondence and annotated music labels.
This semi-supervised approach combines the generalization strengths of self-supervised learning with the domain-specific knowledge of supervised learning.
Moreover, Control-MVR enables explicit control over the relative contribution of each paradigm at inference time, which allows the user to to dynamically prioritize for self-supervised audiovisual patterns or supervised domain knowledge.

We summarize the major contributions of this paper as follows:

\begin{itemize}
    \item To the best of our knowledge, Control-MVR is the first method that combines both self-supervised and supervised training objectives for learning an alignment between music and video.
    
    \item Our model produces a controllable joint embedding space that combines self-supervised and supervised content.
    
    \item Unlike prior work, Control-MVR enables explicit control over the retrieval process at inference time.
\end{itemize}

\section{Approach}
\label{sec:Approach}

\begin{figure}[t]
    \begin{center}
       \includegraphics[width=0.9\linewidth]{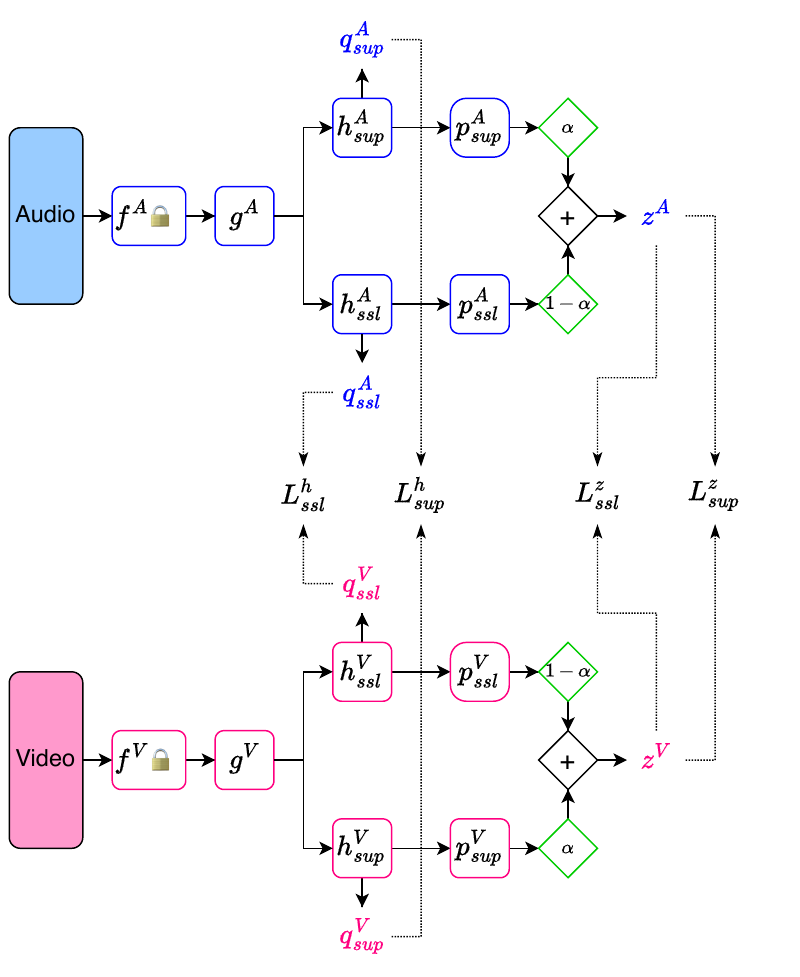}
    \end{center}
    
    \vspace{-2mm}
    \caption{Overview of the semi-supervised \emph{Control-MVR} framework. A dual-branch architecture separately processes music and video, using frozen pre-trained models as well a series of trainable networks. Self-supervised and supervised cross-modal contrastive losses operate on different points in the model architecture. A user-defined weight parameter $ \alpha $ provides explicit control of the output embeddings $ z^A $ and $ z^V $, which are used for music-video retrieval.}
    \vspace{-2mm}

\label{fig:Control-MVR-framework}
\end{figure}

In this section, we describe the training process for the proposed Control-MVR framework (Fig.~\ref{fig:Control-MVR-framework}), as well its application to controllable music-video retrieval.

\subsection{Base Feature Extraction}
\label{subsec:Base Feature Extraction}

Given a music (audio) clip $ x^A $ and a video $ x^V $, we first extract base features $ x_f^A $ and $ x_f^V $ using pretrained audio and video representation models $ f^A(\cdot) $ and $ f^V(\cdot) $, which are kept frozen during training. Each audio/video clip is represented by a single feature vector obtained through temporal aggregation.

\textbf{Audio Features}:
For audio feature extraction, we use MERT \cite{Li-2024-MERT}, an effective music audio representation model trained via large-scale self-supervised learning \cite{Li-2024-MERT}.
We use the \emph{MERT-v1-330M} model, which ingests 24 kHz raw audio and outputs 75 1024-dimensional feature vectors per second.
We compute a learnable weighted average over the outputs of all transformer layers, and then take a global temporal average to produce a single 1024-dimensional feature vector for each audio clip.

\textbf{Video Features}:
For video feature extraction, we use the vision component (\emph{ViT-B/32} architecture) of CLIP \cite{Radford-2021-CLIP}.
We extract a CLIP image feature for each video frame, and then temporally average the features, resulting in a single 512-dimensional feature vector for each video clip.

\subsection{Model Architecture}
\label{subsec:Model Architecture}

As shown in Fig.~\ref{fig:Control-MVR-framework}, the base audio and video features $ x_f^A $ and $ x_f^V $ are separately passed through a dual-branch architecture consisting of multiple trainable networks. The two branches are identical in architecture\footnote{Except for the very first layers of $ g^A(\cdot) $ and $ g^V(\cdot) $, since the base audio and video features have different dimensions.}, but have independent weights. Control-MVR first computes task-specific $ q $ embeddings:
\begin{equation}
\label{eq:framework_q}
    q_{ssl}^M = h_{ssl}^M \left( g^M \left( x_f^M  \right) \right), \quad
    q_{sup}^M = h_{sup}^M \left( g^M \left( x_f^M  \right) \right)
\end{equation}
and then computes output embeddings $ z^A $ and $ z^V $:
\begin{equation}
\label{eq:framework_z}
    z^M = \left( 1 - \alpha \right) \; p_{ssl}^M \left( q_{ssl}^M \right) + \alpha \; p_{sup}^M \left( q_{sup}^M \right)
\end{equation}
where $ M $ represents the audio ($ A $) or video modality ($ V $), $ ssl $ = self-supervised, and $ sup $ = supervised.

The $ g^M(\cdot) $ networks learn representations that are shared across the self-supervised and supervised tasks.
The task-specific networks $ h_{ssl}^M(\cdot) $ and $ h_{sup}^M(\cdot) $ learn task-specific $ q $ embeddings for the self-supervised and supervised tasks.

Finally, the networks $ p_{ssl}^M(\cdot) $ and $ p_{sup}^M(\cdot) $ project $ q_{ssl}^M $ and $ q_{sup}^M $ to the output embedding space where they are combined. As shown in \eqref{eq:framework_z}, $ z^M $ is a linear combination---controlled by a \emph{combination weight} $ \alpha $---of the projected task-specific embeddings.
During training, $ \alpha $ is a hyperparameter between 0 and 1.
The use of $ \alpha $ during inference is explained in subsection \ref{subsec:Controllable Music-Video Retrieval}.

Each of these networks is a multi-layer perceptron, consisting of blocks of \{\emph{linear} $ \rightarrow $ \emph{ReLU} $ \rightarrow $ \emph{dropout}\} layers.

\subsection{Semi-Supervised Contrastive Learning}
\label{subsec:Semi-Supervised Contrastive Learning}

To train Control-MVR, we use semi-supervised contrastive learning between music and video. We use the term \emph{semi-supervised} to indicate that both self-supervised and supervised learning are used. For the remainder of this paper, we use the following notations: $ z_i^M $ = embedding of sample $ i $ of modality $ M $, $ y_i^M $ = label of sample $ i $ of modality $ M $, $ I = \{ 1,...,N \} $ = all indices in a batch, and $ \tau $ = the temperature hyperparameter.



\textbf{Self-Supervised Contrastive Losses}:
We use the cross-modal version of the self-supervised contrastive InfoNCE loss \cite{Oord-2018-InfoNCE}.
The audio-to-video InfoNCE loss is defined as:
\begin{equation}
\label{eq:InfoNCE}
    \ell_{ssl}^{A \rightarrow V} (z) = - \frac{1}{N} \sum_{i=1}^{N}
    log \ \frac{exp \left( z_i^A \cdot z_i^V / \tau \right)} {\sum_{k \in I} exp \left( z_i^A \cdot z_k^V / \tau \right) }
\end{equation}
%
This contrastive loss "pulls together" audio and video embeddings from the same music video and "pushes" apart embeddings from different music videos.
We combine $ \ell_{ssl}^{A \rightarrow V} (z) $ and $ \ell_{ssl}^{V \rightarrow A} (z) $: $ \ell_{ssl} (z) = 0.5 \left( \ell_{ssl}^{A \rightarrow V} (z) + \ell_{ssl}^{V \rightarrow A} (z) \right) $.

\textbf{Supervised Contrastive Losses}:
We use the cross-modal version of the SupCon loss \cite{Khosla-2020-SupCon}, following the approach of \cite{Stewart-2024-EmoCLIM}. 
The audio-to-video SupCon loss is defined as:
\begin{gather}
\label{eq:SupCon}
    \ell_{sup}^{A \rightarrow V} (z) = - \frac{1}{N} \sum_{i=1}^{N} \frac{1}{|P^{A \rightarrow V}(i)|} \sum_{p \in P^{A \rightarrow V}(i)} s(z)
    \\
    \text{where} \ s(z) =  log \ \frac{exp \left( z_i^A \cdot z_p^V / \tau \right) } {\sum_{k \in I} exp \left( z_i^A \cdot z_k^V / \tau \right) }
\end{gather}
$ P^{A \rightarrow V} (i) $ is the set of indices of positive examples $ z_p^V $ for anchor example $ z_i^A $ and is defined as:
\begin{equation}
\label{eq:SupCon_index_sets}
    P^{A \rightarrow V}(i) = \{ p \in I \ | \ y_i^A = y_p^V \}
\end{equation}
This contrastive loss "pulls together" audio and video embeddings with the same label and "pushes" apart embeddings with different labels.
We combine $ \ell_{sup}^{A \rightarrow V} (z) $ and $ \ell_{sup}^{V \rightarrow A} (z) $:
$ \ell_{sup} (z) = 0.5 \left( \ell_{sup}^{A \rightarrow V} (z) + \ell_{sup}^{V \rightarrow A} (z) \right) $.


\textbf{Contrastive Training}:
To train the task-specific $ q $ embeddings, 2 different losses are used: $ L_{ssl}^h = \ell_{ssl} \left( q_{ssl} \right) $ and $ L_{sup}^h = \ell_{sup} \left( q_{sup} \right) $.
$ q_{ssl}^M $ and $ q_{sup}^M $ are trained to contain self-supervised and supervised content, respectively.

To train the output embeddings $ z^M $, 2 different losses are used: $ L_{ssl}^z = \ell_{ssl} \left( z \right) $ and $ L_{sup}^z = \ell_{sup} \left( z \right) $.
$ z^M $ are trained to combine the self-supervised information from $ q_{ssl}^M $ and the supervised information from $ q_{sup}^M $.

In order to train the entire Control-MVR model, multi-task learning is used to balance the multiple objectives previously described. The total loss function that is optimized the sum of all individual losses:
\begin{equation}
\label{eq:total_loss}
    L_{total} = L_{ssl}^z + L_{sup}^z + L_{ssl}^h + L_{sup}^h
\end{equation}

\subsection{Controllable Music-Video Retrieval}
\label{subsec:Controllable Music-Video Retrieval}


As described in subsection \ref{subsec:Model Architecture}, the output embeddings $ z^M $ are linear combinations---controlled by a \emph{combination weight} $ \alpha $---of the projected task-specific embeddings. During inference/retrieval, $ \alpha $ can be adjusted to any value (between 0 and 1) to control the content emphasized in $ z^M $, which in turn controls the retrieval process.

When $ \alpha $ = 0, $ z^M $ maximize \emph{self-supervised} information and minimize \emph{supervised} information.
Hence, the retrieval system maximizes the \emph{self-supervised} similarity
between the query and retrieved item.

When $ \alpha $ = 1, $ z^M $ maximize \emph{supervised} information and minimize \emph{self-supervised} information.
In this case, the retrieval system maximizes the \emph{supervised} similarity
between the query and retrieved item.

By adjusting the value of $ \alpha $ at inference time, the user can dynamically balance the influence of self-supervised learning---which captures broad audiovisual relationships---and supervised learning---which adds more precise domain-specific information.

\section{Experiments and Results}
\label{sec:Experiments and Results}

In all experiments, we use ground-truth music genre labels for the supervised training components of the Control-MVR framework, as well as for supervised retrieval evaluations.

\subsection{Dataset}
\label{subsec:Dataset}

We select a subset of music videos from AudioSet \cite{Gemmeke-2017-AudioSet} that are annotated with precisely one \emph{Music Genre}\footnote{AudioSet ontology: \href{https://research.google.com/audioset/ontology/}{https://research.google.com/audioset/ontology/}} label.
We then condense AudioSet's \emph{Music Genre} classes into a more compact class taxonomy, by grouping similar genres into 11 broader genre categories, as shown in \autoref{table:music_genre_classes}. We do this genre grouping in order to provide cleaner training signals for supervised contrastive learning. Since the SupCon losses treat each class as completely independent, we want to prevent similar genres (e.g., \emph{Blues} and \emph{Funk}) being treated as separate classes.

This process results in a dataset of $106400$ music videos (each with a duration of $10$ seconds) that are annotated with a single music genre label. We split the dataset into $87710$ videos for training, $10000$ for validation, and $8000$ for testing, in a genre-stratified manner.

Since the annotations of AudioSet primarily focus on audio content and effectively ignore visual content, our dataset consists of videos that contain music audio.
Unlike the YouTube-8M dataset \cite{Haija-2016-YouTube8M} used in \cite{Suris-2022-MVPt}, our dataset is not limited to regular music videos (i.e., a video that accompanies music) and includes diverse content such as live concerts, advertisements, and video games. In this paper, we use the term "music video" to refer to any video that contains music audio.

\subsection{Implementation Details}
\label{subsec:Implementation Details}


We use raw audio at a sample rate of $24$ kHz. Since MERT ingests audio inputs of any length \cite{Li-2024-MERT}, we pass the full 10-second audio clips directly into MERT.
We use videos down-sampled to $ 2 $ FPS, and apply CLIP's image pre-processing transforms\footnote{Details can be found at \href{https://github.com/openai/CLIP}{https://github.com/openai/CLIP}.} to each video frame, which include cropping to a size of $ 224 \times 224 $ and normalization.

For supervised contrastive learning, we use a special dataset sampling procedure during training.
To get a single music video, we first randomly select a single genre class, then randomly sample a music video with the selected genre label. In this way, the training batches have a uniform distribution over the genre classes, which helps the supervised contrastive learning process.

We use a dimension of 256 for all joint embedding spaces and use a dropout probability of 0.4 for all networks.
During training, we set $ \alpha $ to 0.5 to equally balance the self-supervised and supervised $ q $ embeddings.
For all contrastive loss functions, we use a temperature of 0.1, which we found to be optimal.
For all training experiments, we use the AdamW \cite{Loshchilov-2019-AdamW} optimizer with a batch size of 1024 and a learning rate of 0.001. We train all models for 50 epochs and keep the model checkpoint with the lowest total validation loss.

\begin{table}[h]
    \caption{Mapping between our music genre classes and AudioSet's original \emph{Music Genre} classes.}
    
    \centering
    \small
    \scalebox{0.82}{
    \begin{tabular}{l | l}
        \toprule
        
        New Genre Class & Original AudioSet Genre Class(es)
        \\
        \midrule
        
        Country & \emph{Country}
        \\
        Classical & \emph{Classical music}
        \\
        Electronic & \emph{Electronic music}
        \\
        \multirow{2}{*}{Non-Western} & \emph{Middle Eastern music}, \emph{Music of Africa}, \emph{Music of Asia}, \\ & \emph{Music of Latin America}, \emph{Traditional music}
        \\
        Hip-Hop & \emph{Hip hop music}
        \\
        Jazz & \emph{Jazz}
        \\
        Pop & \emph{Pop music}
        \\
        Reggae & \emph{Reggae}
        \\
        R\&B & \emph{Blues}, \emph{Disco}, \emph{Funk}, \emph{Rhythm and blues}, \emph{Soul music}
        \\
        Rock & \emph{Rock music}
        \\
        Vocal & \emph{Vocal music}
        \\
        
        \bottomrule
    
    \end{tabular}}
    \vspace{-4mm}

\label{table:music_genre_classes}
\end{table}

\begin{table*}[t]
    \caption{
    Cross-modal retrieval results on a music video test set from AudioSet.
    We report Recall@K (R@K) and Mean Reciprocal Rank (MRR) for self-supervised retrieval, and Precision@K (P@K) and Mean Reciprocal Rank (MRR) for genre-supervised retrieval.
    }
    
    \footnotesize
    \centering
    
    \begin{tabular}{l | l l l | l l l | l l l | l l l}
        \toprule
        
        \multirow{3}{*}{Training Method}{}
        
        & \multicolumn{6}{c |}{Self-Supervised Retrieval} &
        \multicolumn{6}{c}{Genre-Supervised Retrieval}
        \\
        
        & \multicolumn{3}{c |}{Video $ \rightarrow $ Music} &
        \multicolumn{3}{c |}{Music $ \rightarrow $ Video} &
        \multicolumn{3}{c |}{Video $ \rightarrow $ Music} &
        \multicolumn{3}{c}{Music $ \rightarrow $ Video}
        \\
        
        & R@1 & R@10 & MRR &
        R@1 & R@10 & MRR &
        P@1 & P@10 & MRR &
        P@1 & P@10 & MRR
        \\
        \midrule
        
        
        Wav2CLIP \cite{Wu-2022-Wav2CLIP} & \textbf{2.08} & 9.84 & 4.43 & \textbf{1.78} & 9.40 & 4.55 & 27.72 & 25.99 & 41.63 & 27.96 & 26.75 & 42.61
        \\
        
        AudioCLIP \cite{Guzhov-2022-AudioCLIP} & 0.28 & 2.23 & 1.21 & 0.12 & 0.75 & 0.55 & 24.66 & 19.72 & 35.87 & 11.67 & 12.02 & 24.08
        \\
        \midrule
        
        MVPt \cite{Suris-2022-MVPt} & 1.16 & 8.72 & 4.03 & 1.42 & 9.23 & 4.43 & 32.91 & 33.08 & 44.90 & 35.33 & 33.79 & 50.08 \\
        
        Self-Supervised & 1.07 & 8.90 & 4.15 & 1.27 & 9.41 & 4.36 & 34.88 & 33.15 & 46.31 & 34.64 & 33.40 & 49.18
        \\
        
        Supervised & 0.35 & 3.80 & 1.92 & 0.47 & 4.28 & 2.14 & \textbf{46.09} & \textbf{46.07} & 54.03 & \textbf{48.96} & \textbf{50.51} & 58.66
        \\
        
        Semi-Supervised & 0.97 & 7.76 & 3.70 & 1.18 & 8.66 & 4.07 & 44.9 & 43.41 & \textbf{54.70} & 46.09 & 46.15 & 58.04
        \\
        \midrule

        \textbf{Control-MVR}
        \\
        
        \quad $ \alpha = 0.0 $ & 1.35 & 9.78 & 4.45 & 1.60 & 10.41 & 4.81 & 35.95 & 37.13 & 48.87 & 43.65 & 43.12 & 57.86
        \\
        
        \quad $ \alpha = 1.0 $ & 0.62 & 5.13 & 2.52 & 0.73 & 6.08 & 2.86 & 41.3 & 43.3 & 52.59 & 45.67 & 46.74 & 58.48
        \\
        
        \quad Optimal $ \alpha $ & 1.65 & \textbf{10.42} & \textbf{4.83} & 1.76 & \textbf{10.93} & \textbf{5.09} & 42.52 & 43.24 & 53.54 & 47.70 & 46.86 & \textbf{60.32}
        \\
        
        \bottomrule
    
    \end{tabular}

    \medskip
    \vspace{-4mm}

\label{table:retrieval_performance}
\end{table*}

\subsection{Cross-Modal Retrieval}
\label{subsec:Cross-Modal Retrieval}

Following prior works~\cite{Suris-2022-MVPt, Wu-2023-CLAP, Manco-2022-MusCALL, Huang-2022-MuLan, Won-2021-emotion_embedding_spaces, Doh-2023-speech_to_music, Stewart-2024-EmoCLIM}, we evaluate Control-MVR with cross-modal retrieval tasks, using the held-out test set.

\textbf{Experimental Setup}:
Given a video/music query, we retrieve the $ K $ most similar music/video clips in the test set, ranked by the cosine similarity between the output audio and video embeddings
$ z^A $ and $ z^V $.
We run two different types of retrieval experiments: \emph{self-supervised retrieval} and \emph{genre-supervised retrieval}.

For self-supervised retrieval,
a retrieval is successful if the retrieved music/video clip belongs to the exact same music video as the query. Following \cite{Suris-2022-MVPt}, we use an evaluation set size of 2000 music videos.
We divide the full test set of 8000 music videos into 4 disjoint subsets of size 2000, and then average the results of the subsets.
We report Recall@K (R@K) and Mean Reciprocal Rank (MRR) scores.

For genre-supervised retrieval,
a retrieval is successful if the retrieved music/video clip has the same genre label as the query. We directly use the full 8000-video test set, since supervised retrieval results do not depend on the evaluation set size. We report Precision@K (P@K) and Mean Reciprocal Rank (MRR) scores. Following previous works \cite{Won-2021-emotion_embedding_spaces, Doh-2023-speech_to_music, Stewart-2024-EmoCLIM}, we macro-average retrieval metrics across all genre classes
to avoid potential bias caused by class imbalances.

\textbf{Baseline Approaches}:
We compare against three other works: Wav2CLIP \cite{Wu-2022-Wav2CLIP}, AudioCLIP \cite{Guzhov-2022-AudioCLIP}, and MVPt \cite{Suris-2022-MVPt}. We use open-source implementations of Wav2CLIP and AudioCLIP. Since the model weights of MVPt are not available, we replicated and trained the model on our dataset---but modified the input video/music lengths to be compatible with AudioSet's 10-second clips.\footnote{We used 10 1.0-second segments to represent each music video.}
We do not conduct a direct comparison with the recent work on video-to-music retrieval presented in \cite{McKee-2023-language_guided_mvr}, as their approach incorporates language in addition to audio and video inputs, which differs from our task. Additionally, their model weights are not publicly available.

We also train several other baseline approaches: a fully self-supervised model, a fully supervised model, and a simpler semi-supervised model.
These baselines use a simplified version of the Control-MVR architecture in Fig.~\ref{fig:Control-MVR-framework}, where only the $ g^A(\cdot) $ and $ g^V(\cdot) $ networks are trained,
and their outputs are used as the final output embeddings.
For the self-supervised/supervised baselines, a single self-supervised/supervised contrastive loss is used to train the model. For the semi-supervised baseline, an equally-weighted combination of self-supervised and supervised contrastive losses is used.

\textbf{Results}:
\autoref{table:retrieval_performance} lists the self-supervised and genre-supervised cross-modal retrieval results.
All metrics are shown as percentages.
The rows are divided into 3 sections: 1) the first section contain baseline methods (Wav2CLIP and AudioCLIP) that were trained on different datasets, 2) the second section contains baseline methods trained on our dataset, 3) and the third section is for Control-MVR.

For Control-MVR, we show results with 3 different inference-time $ \alpha $ values: 0.0, 1.0, and an optimal value.
The optimal value is the $ \alpha $ value in $ \{0.0, 0.1,...,0.9, 1.0 \} $ that results in the best Recall@10 and Precision@10 score on the validation set, for self-supervised and genre-supervised retrieval, respectively.
The optimal $ \alpha $ value is 0.4 for self-supervised retrieval and 0.8 for genre-supervised retrieval.

For self-supervised retrieval, Control-MVR (with optimal $ \alpha $) noticeably outperforms all baselines except for Wav2CLIP, which it marginally outperforms (except for Recall@1).
However, we note that Wav2CLIP is trained on a larger dataset \cite{Chen-2020-VGGSound} and also trains the entire audio encoder, unlike Control-MVR.
Importantly, Control-MVR outperforms the fully self-supervised baseline,
which demonstrates that genre-label supervision can complement ordinary self-supervised learning.
Furthermore, the semi-supervised baseline does not outperform the self-supervised baseline, which suggests that the more complex Control-MVR model architecture is necessary to effectively combine the self-supervised and genre-supervised objectives.

For genre-supervised retrieval, Control-MVR significantly outperforms all self-supervised baselines (including Wav2CLIP and AudioCLIP), but slightly underperforms the fully-supervised baseline.
However, the supervised baseline overfits to the music genre labels,
which is clear from its weak self-supervised retrieval performance.


Control-MVR shows SOTA or near-SOTA performance for both self-supervised and genre-supervised retrieval evaluations, showing an advantage over the other baselines, which are competitive at only a single type of evaluation. Moreover, Control-MVR adds retrieval controllability, which is explored further in the next section.

\subsection{Retrieval Controllability}
\label{subsec:Retrieval Controllability}

\begin{figure}[t] {
    \centering
    
    \begin{tikzpicture}
        
        \tikzstyle{label_text} = [text centered]
        
        \node[label_text] at (0.0\linewidth, 0.22\linewidth) {a) Self-Supervised Retrieval};
        \node[draw=none,fill=none] at (0.0\linewidth, 0.0\linewidth){\includegraphics[width=0.47\linewidth]{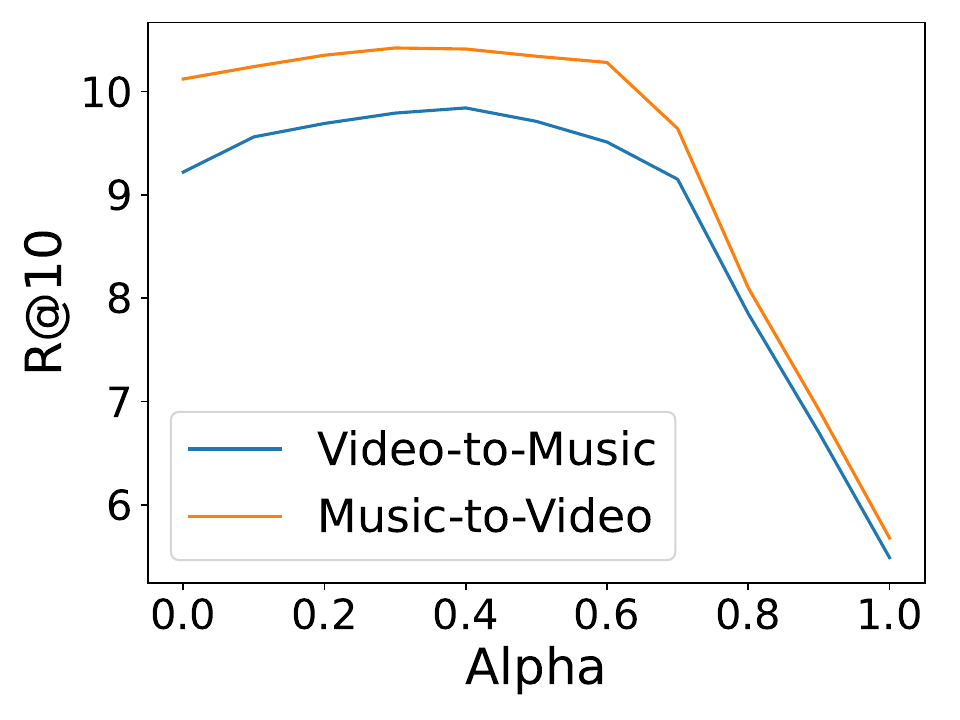}};
        
        \node[label_text] at (0.5\linewidth, 0.22\linewidth) {b) Genre-Supervised Retrieval};
        \node[draw=none,fill=none] at (0.5\linewidth, 0.0\linewidth){\includegraphics[width=0.47\linewidth]{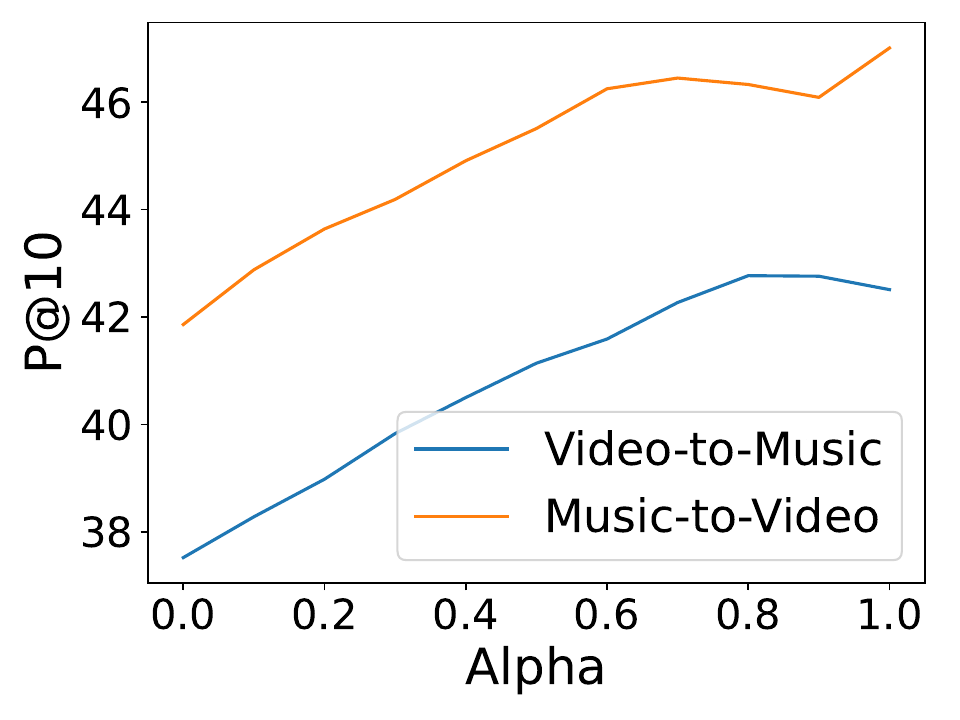}};
    
    \end{tikzpicture}
    
    \caption{Control-MVR enables explicit control over the retrieval process at inference time. a) Decreasing $ \alpha $ increases self-supervised content in the output embeddings, which in turn improves self-supervised retrieval performance. b) Conversely, increasing $ \alpha $ increases genre-supervised content in the output embeddings, which improves genre-supervised retrieval performance.}
    \vspace{-2mm}
\label{fig:controllability_plots}
}
\end{figure}

To explore the controllability of Control-MVR, we vary the value of $ \alpha $ (at inference time) from 0.0 to 1.0 in steps of 0.1, and repeat the retrieval experiments for each $ \alpha $ value. As shown in Fig.~\ref{fig:controllability_plots}
we plot Recall@10 (R@10) vs. $ \alpha $ for self-supervised retrieval and Precision@10 (P@10) vs. $ \alpha $ for genre-supervised retrieval.

For self-supervised retrieval,
Recall@10 slightly increases then sharply decreases.
Since a larger $ \alpha $ value results in less self-supervised content in the output embeddings, we can generally expect an inverse relationship, which is clearly shown for larger $ \alpha $ values. But interestingly, the peak Recall@10 scores occur at $ \alpha $ values of around $ 0.4 $, which suggests that partially including genre-supervised content actually improves the self-supervised retrieval task.

For genre-supervised retrieval,
Precision@10 generally increases as $ \alpha $ increases, which is expected since a larger $ \alpha $ value results in more genre-supervised content in the output embeddings. Analogous to the self-supervised retrieval plot, the peak Precision@10 scores occur at $ \alpha $ values of around $ 0.8 $, which indicates that some amount of self-supervised content is beneficial for genre-supervised retrieval.

\section{Conclusion}
\label{sec:Conclusion}

In this paper, we introduced Control-MVR, a novel framework for controllable music-video retrieval. Our approach combines both self-supervised and supervised contrastive training objectives to learn a controllable joint embedding space between music audio and video. We demonstrated the effectiveness of Control-MVR by training a model on a music video dataset annotated with music genre labels.
Our model outperformed all baseline methods on self-supervised retrieval tasks, while being comparable to a fully supervised baseline on genre-supervised retrieval tasks.
Through retrieval controllability experiments, we also showed that Control-MVR can effectively guide the retrieval process at inference time.

For future work, we plan to extend this framework by incorporating additional music video labels (e.g., emotion) in alongside to genre, enabling the model to balance multiple music attributes (e.g., genre, emotion) during retrieval.
Furthermore, we aim to explore the integration of language-guidance into the Control-MVR framework.
Our results showed that using semi-supervised learning can facilitate a controllable music-video retrieval system with strong performance, and we hope that our work can help motivate further research in controllability of cross-modal retrieval systems.

\clearpage

\bibliographystyle{IEEEtran}
\bibliography{references}

\end{document}